\documentclass[aps,prd,epsf,amsmath,amssymb, 11pt]{revtex4}
\usepackage{dcolumn}
\usepackage{bm}

\newcommand{\be}{\begin{equation}} 
\newcommand{\ee}{\end{equation}}
\newcommand{\ba}{\begin{eqnarray}}
\newcommand{\ea}{\end{eqnarray}}
\newcommand{\ban}{\begin{eqnarray*}}
\newcommand{\ean}{\end{eqnarray*}}

\usepackage{color,graphicx}

\begin{document}
\title{Spacetime Singularities and Cosmic Censorship}

\author{Pankaj S. Joshi}
\affiliation{Tata Institute for Fundamental Research\\
Homi Bhabha Road,  Mumbai 400005, India}

\begin{abstract} We present here a brief review
and discussion on recent developments in the theory of
spacetime singularities. After mentioning some key 
motivations on the main ideas and concepts involved,
we take the approach that the singularities will be
eventually resolved by the quantum gravity effects.
Some consequences are indicated when such singularities 
are visible to far away observers in the universe.

\end{abstract}


\maketitle

\section{Introduction}

Physical phenomena in the universe take place in the arena of space, 
and evolve in time. The known laws of physics describe and govern 
these happenings that occur in nature. But then what are space and time, 
what are their interconnections if any, and how the space and time 
themselves originate, or whether each of these is actually infinite 
and endless? These are some of the most profound questions that have 
exercised greatest of minds in science and philosophy over past 
many centuries.

The best possible scientific theory that we have today, that governs 
the universe of space and time is the general theory of relativity. In 
relativity, the space and time are no longer separate and independent of 
each other, but they are intertwined with each other and the actual 
physical measurements of these quantities are always mutually related. 
While general relativity was originally developed to describe the force 
of gravitation, it has provided us with some of the most intriguing 
insights on the nature of space and time, their inter-connections, and 
the possible origins of the time and space itself. General relativity 
suggests not to view gravity as a `force' in the usual sense, but describes 
it as the curvature and geometry of the `space-time continuum', which 
is our universe.

General relativity describes the interply of the space-time curvature, 
and the matter within it that generates such a curvature, just as a metal 
ball placed on a rubber sheet curves it. It is a classical theory that
governs the universe in its large scale structure. Given any physical system
governed mainly by gravity, such as a large collection of galaxies, or a
massive star that it close to the end of its evolution having burnt 
all its nuclear fuel, the Einstein equations govern the future evolution of
such a system in time. Thus we can ask the questions such as how the 
universe of galaxies that is continuously expanding will evolve in future,
and whether it will continue to expand or will it start shrinking at some 
point in time in future. Or one could ask, what will be the final end point 
of evolution of a massive star that has started contracting under the 
force of its own gravity when its internal fuel is exhausted.

One of the most remarkable predictions of general relativity, developed
during the 1960s and early 1970s has been that, dynamical evolution of
matter fields in a space-time generically produces a space-time singularity.
Such a singularity is where the densities, space-time curvatures and all 
other physical quantities blow up and grow arbitrarily large, and thus all 
known physical laws no longer hold there. In that sense, the singularity 
is the end (or beginning) of the space and time themselves.

The singularity theorems developed by Roger Penrose, Stephan Hawking
and Robert Geroch show that the evolution of matter fields in a 
spacetime generically yields such a singularity, provided reasonable physical 
conditions are satisfied such as the causality ensuring that you do not return
to your own past, a suitable energy condition ensuring the positivity of 
energy density, and formation of what are called `trapped surfaces' in a 
space-time that indicate and characterize that the gravitational field is 
sufficiently 
strong. The space-time singularities develop in cosmology, where they signal 
the beginning of time, and in gravitational collapse of massive stars, 
which is an issue of great interest in gravitation physics today that has 
been investigated in much detail in recent years in the Einstein theory.

We outline here some aspects of space-time singularities that occur 
in cosmology and in gravitational collapse. The singularity theorems predicting 
the occurrence of singularities allow the singularities of gravitational 
collapse to be either visible to external observers or covered by an event 
horizon of gravity. Some consequences of this fact are indicated. The role 
of space-time singularities as an inevitable feature of Einstein's theory of 
gravity has became clear now as signalling the situations where the 
gravitational field becomes ultra-strong and grows without any upper bound.
Close to the singularity is the regime of strong gravity fields, where 
general relativity comes into its own to imply most interesting 
physical consequences.

\section{The occurrence of Singularities}

We discuss now the occurrence of space-time singularities in some
detail within a general spacetime framework. The basic ideas involved in 
the singularity theorems are indicated, and what these theorems  
do {\it not} imply is pointed out.

We observe the universe today to the very far depths in space 
and time through the telescopes that observe the objects which are 
billions of light years away. One could look deep into space, let us
say in diametrically opposite directions. Then the regions with 
extremely distant galaxies are seen in each of these directions. It is 
most interesting to observe that these regions have actually quite similar 
properties in terms of their appearance and there is a homogeneity seen 
in the spatial distribution of the far away galaxies. The universe also 
looks similar in different directions, thus exhibiting an isotropy. 
These regions are, however, so far away from each other that they have 
had no time to interact mutually. That is because, general relativity 
equations imply that a homogeneous and isotropic universe had a finite 
age in the past when the energy density of matter is positive. The age 
of the universe since such a big bang that indicated the origin of 
both space and time has not been actually large enough for any such 
interactions to have taken place in the past. Thus, within the big bang 
framework of cosmology, a very relevant question arises: How come these 
regions have such similar properties? This is one of the major 
puzzles of modern cosmology today.

This observed homogeneity and isotropy of the universe at large 
enough scales can be modelled by the so called Friedmann-Robertson-Walker 
geometry. The metric describing the geometry of the corresponding 
space-time universe is given by
\begin{equation}
ds^2 = - dt^2 + R^2(t) \big[ {dr^2\over (1-kr^2)} + r^2 d{\Omega}^2 \big]. 
\end{equation}
Here $d\Omega^2 = d\theta^2 + sin^2\theta d\phi^2$  is the
metric on a two dimensional sphere and the universe is assumed to be 
spherically symmetric here. The additional assumption here is that the 
matter content of the space-time is homogeneous and isotropic, to represent
these observed features of the universe. The scale factor $R(t)$ increases
in time so as to model the observed expansion of the universe where
galaxies recede from each other in space. Thus, the matter density is 
the same everywhere in the universe at any given epoch of time, and also 
the visual appearance of universe looks the same in all directions.

The galaxies here are represented by point-like objects which form
`dust particles' of this universe, which is the matter content of the 
space-time. Combining these geometrical features with the Einstein equations 
and solving the same, one is led to the Friedmann solution yielding a 
description of the dynamical evolution of the universe and the matter
within it. The picture obtained from such an evolution implies that the 
universe must have had a beginning at a finite time in the past. This is the 
epoch of the so called big bang singularity. The matter density as well as 
the curvatures of spacetime diverge in the limit of approaching this 
cosmological singularity. This is an epoch where all non-spacelike geodesics
that represent the trajectories of the photons and the material particles
come to an end and these are `incomplete' at a point in the past where 
the space-time comes to an end.

A similar occurrence and formation of a space-time singularity 
takes place when a massive star collapses freely under the force of its 
own gravity when it has exhausted its internal fuel which made it shine
earlier. If the mass of the star is small enough, it can stabilize as a 
white dwarf or a neutron star at the end of its life cycle, which 
would then be the natural endstate of its evolution. However, in case 
the mass is much larger, say of the order of tens of solar masses, a continual 
gravitational collapse of the star is inevitable once there are no internal 
pressures left to sustain the star. This is because no known forces can
then stabilize such a star. Such a scenario was considered and 
modelled using general relativity by R. Oppenheimer and H. Snyder in 1939, 
and by B. Datt in 1938, when they considered a collapsing spherical cloud 
of dust. Again, according to the equations of general relativity, a 
space-time singularity of infinite density and curvature forms at the 
center of the collapsing cloud. We shall discuss gravitational 
collapse in some more detail in the next section.

Such space-time singularities were discovered in the context of 
specific models of universe or of a collapsing massive star, such as
those discussed above, and after their discovery these were debated 
extensively by the gravitation theorists in the 1940s and 1950s. 
An important key question that was persistently asked at this juncture 
in this connection was the following: Why should these models be taken 
seriously at all, when they were so special because they assumed so 
many symmetries of space-time? As a result, perhaps such a space-time 
singularity was arising due to these special symmetry assumptions, and may 
be it could occur in such special circumstances only as described and 
assumed by these models, but possibly it would not actually develop  
in the actual physical reality which is our universe. In other words, 
these singularities could be just some isolated examples occurring in 
some special models and manifestation of the symmetry assumptions made. 
After all, the Einstein equations
\begin{equation}
R_{ab} -  {1 \over 2} R g_{ab} = 8\pi T_{ab},
\end{equation}
governing the ever present force of gravity, are a complex system of
second-order, non-linear, partial differential equations, which admit an
infinite space of solutions. The models discussed above are only
special cases and isolated examples in this full space of solutions.

Therefore, the main issue was the absence of any general enough
proof that such space-time singularities would always occur in a general 
enough gravitational collapse depicting actual physical systems when a 
massive star dies, or in a generic enough cosmological scenario. 
In fact, there was a widespread belief in the 1940s and 1950s that such 
singularities would be simply removed and go away both from stellar collapse 
and from the cosmological considerations of the universe (which are two 
very important physical situations), once assumptions such as the dust 
form of matter, the spherical symmetry of the model, and such others 
were relaxed and when more general solutions to the Einstein equations 
were found and considered.

This is where the work by A. K. Raychaudhuri in 1955, on the 
gravitational focusing of matter and light in a space-time universe 
became relevant. This was used by R. Penrose, S. W. Hawking, and R. Geroch, 
who analyzed the causal structure and global properties of a fully general 
spacetime, and who then combined it with the considerations on 
gravitational focussing effect as developed by Raychaudhuri. The culmination 
of these efforts was the singularity theorems in general relativity, 
which showed that space-time singularities, such as those depicted 
in the examples of gravitational collapse and cosmology we discussed above, 
in fact manifested themselves in a rather large class of space-time 
universes under quite general physical conditions.

The Raychaudhuri equation played a central role in the analysis of 
space-time singularities in general relativity. Prior to the use of this 
equation to analyse collapsing and cosmological situations for the 
occurrence of singularities 
\cite{wald}
\cite{joshi2}, 
most works on related issues had considered only rather special cases 
with many symmetry conditions assumed on the underlying spacetime. 
But with the help of this equation these aspects could be discussed 
within the framework of a general spacetime without any symmetry conditions. 
This was in terms of the overall behaviour of the congruences of trajectories 
of material particles and photons propagating and evolving dynamically. 
This analysis of the congruences of non-spacelike curves which represent 
either material particles or light rays, showed how gravitational focusing 
took place in the universe giving rise to what are called caustics where
nearby trajectories intersect due to gravitational focussing.

Before general singularity theorems could be constructed, however, 
another important mathematical input was needed in addition to the Raychaudhuri 
equation. This was the analysis of the causality structure and general 
global properties of a spacetime manifold. This particular development 
took place mainly in the late 1960s (for a detailed
discussion, see e.g. 
\cite{geroch}).   
The singularity theorems then combined these two important features, 
namely the gravitational focusing effects due to matter and energy content 
of the space-time and the causal structure constraints which followed 
from the global spacetime properties, to obtain the existence of 
space-time singularities in the form of geodesic incompleteness in the  
space-time.

As we discussed earlier, the main question here was that of 
genericity of the space-time singularities, either in cosmology or in 
collapse situations. The singularity theorems, while proving the existence 
of singularities in a generic manner, by themselves provide no information 
either on the structure and properties of such singularities, or on 
the growth of curvature and densities in their vicinity.

There are several singularity theorems available which establish 
the non-spacelike geodesic incompleteness for a spacetime under different 
sets of physical conditions. Each of these may be more relevant to 
one or the other specific physical situation, and may be applicable to 
different physical systems such as stellar collapse or the universe as 
a whole. However, the most general of these is the Hawking-Penrose theorem, 
which is applicable to both the collapse situation and the cosmological 
scenario. To outline briefly the basic idea and the chain of logic 
behind the same, firstly, using causal structure analysis it is shown that 
between certain pairs of events in space-time there must exist 
timelike geodesics curves of maximal length. However, both from 
causal structure analysis and from the global properties of a general 
space-time manifold (which is assumed to satisfy a specific energy 
condition), it follows that a causal geodesic curve, which is complete 
in regard to both the future and past, must contain caustics where 
nearby null or timelike geodesics must intersect. One is then led to a 
contradiction, because the maximal geodesic curves mentioned above 
are not allowed to contain any such conjugate points, the existence 
of which would be against their maximality. Thus the space-time itself 
must have non-spacelike geodesic incompleteness.

Such theorems do assume some physical reasonability conditions. 
First such condition assumed on the space-time is an energy condition. 
All classical fields have been observed to satisfy a suitable 
positivity of energy density requirement, and therefore this may be 
considered to be quite reasonable. The second condition is a statement 
that all non-spacelike trajectories do encounter some non-zero matter 
or stress-energy density somewhere during their entire path. This is 
called the genericity condition. The third is a global causality 
requirement to the effect that there are no closed timelike curves in 
the spacetime. Finally, there is a condition that relates to either 
a gravitational collapse situation or to gravitational focussing 
within a cosmological framework, which considers how the congruences 
of non-spacelike curves in a space-time expand or converge. If 
these conditions are satisfied then the theorem goes through,
proving the existence of space-time singularities within a general
space-time scenario without special symmetry conditions.

\section{Gravitational Collapse}

Another important physical situation where space-time singularities
do occur is the gravitational collapse of a massive star. In fact, it was
pointed out by S. Chandrasekhar in 1935 that, "..the life history of
a star of small mass must be essentially different from that of a star of
large mass... A small mass star passes into White-dwarf stage... A star of
large mass {\it cannot} pass into this stage and one is left speculating
on other possibilities."

While we have pointed out above that space-time singularities must
develop in gravitational collapse, the most important question that 
this situation gives rise to is: {\it What is the final fate of a massive 
star when it undergoes a continual gravitational collapse}? This has been 
one of the most important key problems in astronomy and astrophysics for 
past many decades. If the star is sufficiently massive, beyond the 
white dwarf or neutron star mass limits, then a continued gravitational 
collapse must ensue when the star has exhausted its nuclear fuel.

What are then the {\it possible end states} of such a {\it continued 
gravitational collapse} is the issue to be resolved. To answer this 
question, one must study dynamical 
collapse scenarios within the framework of a gravitation theory 
such as the Einstein theory. We now outline some recent 
developments in past couple of decades in this area on the final state 
of a gravitationally collapsing massive matter cloud. We point out 
how the black hole and naked singularity end states arise naturally 
and generically as spherical collapse final states. We see that it 
is the geometry of trapped surfaces in the space-time that governs this 
phenomena.

It was conjectured by Penrose in 1969, that the ultra-dense 
regions forming in gravitational collapse, that is the space-time 
singularities where the physical quantities such as densities and 
curvatures are having extreme values, must be hidden within the event 
horizon of gravity. That is, the collapse must end in a black hole. 
This is called the `cosmic censorship conjecture'. There is, however, 
no proof or any suitable mathematical formulation available for 
the same as of today despite many attempts.

If the gravitational collapse always produces a black hole, then 
that provides a very strong foundation for the theory as well as the
astrophysical applications of black holes. On the other hand, if 
collapse produces visible ultra-strong gravity regions, or naked 
singularities, then the physical processes in these super-strong 
gravity regions can propagate, in principle, to external observers in 
the universe, thus giving rise to very interesting physical 
consequences.

Under the situation, very many researchers have made extensive 
studies of various dynamical collapse models, mainly spherically symmetric, 
over past couple of decades, to investigate the final outcome of a continual 
gravitational collapse. When no proof, or even a suitable mathematical 
formulation of censorship conjecture is available, it is only such 
studies that can throw light on this issue. The generic conclusion that
follows is: Either a {\it black hole} or a {\it naked singularity}
develops as end product of collapse, depending on the initial data for 
the collapsing matter cloud (for example, the initial density, 
pressures, and velocity profiles for the collapsing shells of matter), 
from which the collapse develops, and the nature of dynamical evolutions
as permitted by Einstein equations.

While extensive study is made of astrophysics of black holes, for 
the visible singularities we may still want to inquire into questions 
such as: Are naked singularities of gravitational collapse generic, or 
What are the physical factors that cause a naked singularity, rather 
than a black hole forming as collapse end state. That is, one may 
wish to understand in a better way the naked singularity formation in 
gravitational collapse. Basically, it turns out that the black hole 
or naked singularity phases of collapse are determined by the
geometry of the trapped surfaces that develop as the collapse evolves.

What governs the geometry of the trapped surfaces, or the 
formation or otherwise of the naked singularities in the spacetime?
We can ask in other words, what is it that causes the naked singularity 
to develop rather than a black hole as collapse final state? 
It turns out that physical agencies such as inhomogeneities in matter 
profiles play an important role to distort the trapped surface 
geometry to delay the trapped surface formation during the collapse, 
thus giving rise to a naked singularity.

When the collapsing dust matter is homogeneous, the final outcome 
of collapse is a black hole and the singularity is hidden within 
the horizon. But if the collapsing cloud has a density higher at 
the center, then the trapped surfaces are delayed and the outcome is 
a naked singularity from which the light or matter particles 
can escape away
(see Fig.1 and Fig.2).

\begin{figure}
\begin{center}
\includegraphics[width=10cm]{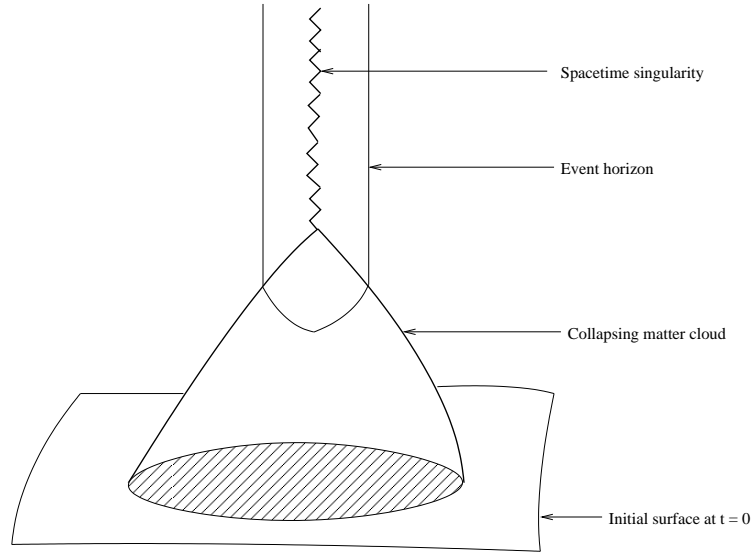}
\caption{\label{Fg1}
The space-time singularity developing in collapse is 
hidden within an event horizon in the case of gravitational collapse 
of a homogeneous density dust cloud.
}
\end{center}
\end{figure}

\begin{figure}
\begin{center}
\includegraphics[width=10cm]{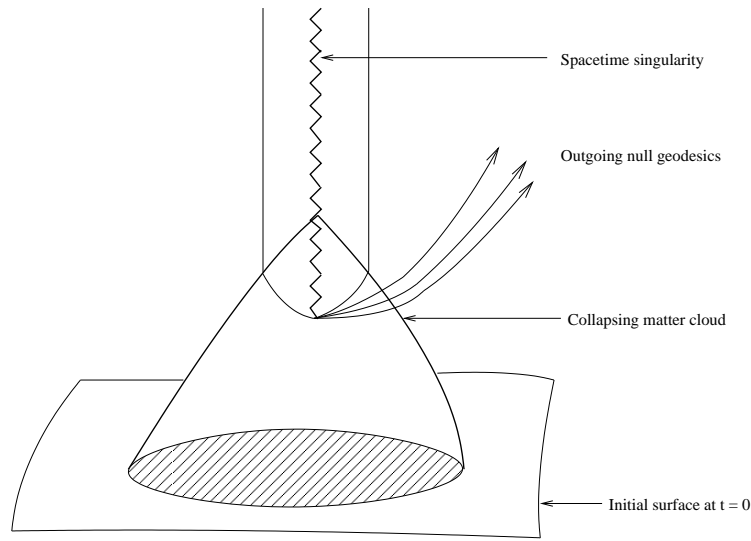}
\caption{\label{Fg2}
When the density of collapsing cloud is higher at the 
center, light rays or particle trajectories can reach the external 
observer from the vicinity of the singularity.
}
\end{center}
\end{figure}

This, in a way, provides the physical understanding of the 
phenomena of black hole and visible singularities occurring as 
end states for gravitational collapse.

What would be the outcome when the collapse is non-spherical? 
There are some examples which indicate the outcome to be somewhat similar 
in nature, but the evidence in this case is limited so far. The main 
difficulty is the complexity of the Einstein system of differential 
equations. It will be necessary to understand non-spherical collapse
better before we can decide on the genericity aspect of the visible
singularities.

There are several quite interesting questions which are under
active investigation at the moment. For example, could naked
singularities generate bursts of gravity waves? What kind of quantum
effects will take place near a visible singularity? Many of these issues
would have interesting physical implications. It appears likely from 
the current investigations that the astrophysical phenomena such
as the Gamma Rays Bursts will have a strong connection to the physics 
and dynamics of gravitational collapse of massive stars.

The above discussion points to a wide variety of circumstances
under which singularities develop in general relativistic cosmologies
and in many gravitational collapse processes. Singularity theorems imply
the existence of vast classes of solutions to the Einstein equations that 
must contain spacetime singularities, as characterized by the conditions
of these theorems, and of which the big bang singularity is one 
example. These theorems therefore imply that singularities must occur 
in Einstein's theory quite generically, that is, 
under rather general physically reasonable conditions on the underlying 
spacetime. Historically, this implication considerably strengthened our 
confidence in the big bang model which is used extensively 
in cosmology today.

While singularity theorems tell mainly on the existence part,
what we really need is more information on the structure of the 
singularities in terms of their visibility or otherwise, curvature strengths 
and other such aspects. What is therefore called for is a detailed 
investigation of the dynamics of gravitational collapse within the framework 
of Einstein's theory.

In such a context, discussion of the gravitational collapse for specific
models in general relativity can turn out to be of great help. One such
model is that given by the Vaidya metric, which was originally developed
by P. C. Vaidya in 1941 in the context of modelling a radiating star
in general relativity. One can use this metric to study collapse of 
radiation shells within a Vaidya geometry, and it provided a great deal
of information on the black hole and naked singularity formation
in such collapse geometries 
\cite{joshi1}, \cite{joshi2}).

\section{Singularities and Quantum Gravity}

Though general relativity deals with matter in the space-time as a 
purely classical entity that generates the curvature of space-time, we 
know that actually the matter and particles, and their interactions,
obey and are governed by the laws of quantum theory. On very large 
scale in the universe, and at relatively lower matter densities, it may
be possible to ignore the intrinsic quantum nature of the matter.
In that case, general relativity provides us with fairly accurate 
predictions on the evolution of the universe.

The occurrence of singularities, however, offers us the regime 
where the matter densities, space-time curvatures, and gravity are 
all indeed extreme, and where the quantum gravity effects would be 
certainly important. As of today, we do not yet have a combined theory 
governing in a unified manner the forces operating within atom and 
at nuclear densities, and the force of gravity. Therefore a study of 
physical processes occurring in the vicinity of the space-time singularity 
would possibly offer a unique opportunity to study the physics
of the gravity and the quantum together, and could possibly lead to 
a unified theory of all forces of nature. This is the cherished dream of 
the physicists which is to create a quantum theory of gravity. It is 
for this reason that the study of space-time singularities occupies 
such a central place in fundamental physics today.

It is possible that such singularities represent the incompleteness
of the theory of general relativity itself. Further, they may be 
resolved or avoided when quantum effects near the same are included in a 
more complete theory of quantum gravity. Nevertheless, there is a 
key point here. Even if the final singularity is dissolved by quantum 
gravity, what is really important is the inevitable occurrence of an 
ultra-strong gravity region, close and in the vicinity to the location 
of the classical singularity, either in cosmology or in dynamical processes 
involved in gravitational collapse. Such processes must affect the 
physics of the universe. An example of such a situation is the big bang 
singularity of cosmology. Even though such  singularities may be 
possibly resolved through either quantum gravity effects, or due to 
features such as chaotic initial conditions, the effects of the super 
ultra-dense region of gravity that existed near the big bang epoch profoundly 
influence the physics and subsequent evolution of the universe.  
Similarly, we have in gravitational collapse of a massive star,  
the occurrence of singularities which are either visible to external 
observers or hidden behind the event horizon giving rise to a black hole. 
In either case again, the important issue is how the inevitably 
occurring super ultra-dense region would influence the physics outside.

\begin{figure}
\begin{center}
\includegraphics[width=10cm]{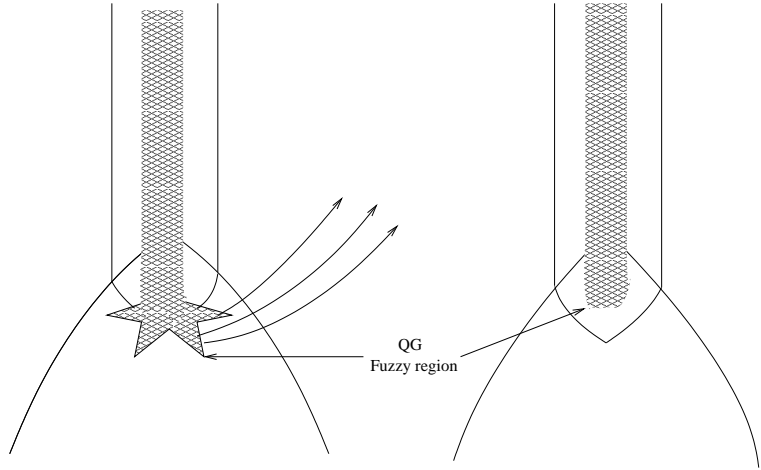}
\caption{\label{Fg3} Eventhough the singularity may resolve by the quantum 
gravity effects, the physical processes in the ultra-strong gravity 
regions may be seen by the observers far away in the universe.
{\it (Figure courtesy: Ref 3 below of the PSJ book.)}
}
\end{center}
\end{figure}

We discussed here some aspects of space-time singularities. 
It is seen that in Einstein gravity they occur generically, whether 
covered within event horizons or as visible to external observers. 
If a future quantum theory of gravity resolves the final singularity 
of collapse or the initial one in cosmology, what is really interesting 
physically is the occurrence of regions of ultra-strong gravity and 
space-time curvatures, that develop as the result of the dynamical 
gravitational processes.

The following physical picture then emerges. 
Dynamical gravitational processes proceed and evolve to create 
ultra-strong gravity regions in the universe. Once these form, strong 
curvature and quantum effects both come into their own in these regions. 
Quantum gravity then takes over and may resolve the final singularity. 
Particularly interesting is the case when the singularities
of collapse are visible. In such a situation, quantum gravity effects,
taking place in those ultra-strong gravity regions, will in principle be 
accessible and observable to external observers 
(see Fig.3).
The consequences of such a scenario would be surely intriguing.

\end{document}